\documentstyle[12pt,epsfig]{article}

\newlength{\dinwidth}
\newlength{\dinmargin}
\def\numu{\nu_\mu}
\def\nutau{\nu_\tau}

\def\numunutau{\nu_\mu\rightarrow \nu_\tau}

\def\Journal#1#2#3#4{{#1} {\bf #2}, #3 (#4)}

\def\NIMA{{\em Nucl. Instrum. Methods} A}

\def\PLB{{\em Phys. Lett.}  B}
\def\PRL{\em Phys. Rev. Lett.}

\begin{document}
\pagestyle{empty}

\vspace*{3cm}
\begin{center}
{\Large \bf  Measurement of Neutrino Oscillations} \\
\vspace{0.3cm}
{\Large \bf by means of a High-Density Detector} \\
\vspace{0.3cm}
{\Large \bf on the Atmospheric Neutrino Beam} \\
\end{center}
\vspace{3cm}
\begin{center}
{\large
M.~Aglietta$^{\rm{a}}$, 
M.~Ambrosio$^{\rm{b}}$,
E.~Aprile$^{\rm{c}}$,
G.~Bologna$^{\rm{a,d}}$, \\
M.~Bonesini$^{\rm{e}}$,
G.~Bencivenni$^{\rm{d}}$,
M.~Calvi$^{\rm{e}}$,
A.~Castellina$^{\rm{f}}$, \\
A.~Curioni$^{\rm{c}}$,  
W.~Fulgione$^{\rm{f}}$, 
P.L.~Ghia$^{\rm{f}}$,
C.~Gustavino$^{\rm{g}}$, \\
R.P.~Kokoulin$^{\rm{h}}$,
G.~Mannocchi$^{\rm{d,f}}$, 
F.~Murtas$^{\rm{d}}$, 
G.P.~Murtas$^{\rm{d}}$, \\ 
P.~Negri$^{\rm{e}}$, 
M.~Paganoni$^{\rm{e}}$, 
L.~Periale$^{\rm{f}}$,
A.A.~Petrukhin$^{\rm{h}}$, \\
P.~Picchi$^{\rm{d,f,a}}$, 
A.~Pullia$^{\rm{e}}$, 
S.~Ragazzi$^{\rm{e}}$, 
N.~Redaelli$^{\rm{e}}$, \\ 
L.~Satta$^{\rm{d}}$, 
T.~Tabarelli de Fatis$^{\rm{e}}$, 
F.~Terranova$^{\rm{e}}$,  \\
A.~Tonazzo$^{\rm{e}}$,
G.~Trinchero$^{\rm{f}}$,
P.~Vallania$^{\rm{f}}$,
B.~Villone$^{\rm{f}}$
}
\end{center}
\vspace{3cm}
\begin{center}
{
$^{\rm{a}}$ {\it Dipartimento di Fisica, Universit\`a di Torino, 
                 Torino, Italy } \\
$^{\rm{b}}$ {\it INFN, Sezione di Napoli,  Napoli, Italy } \\
$^{\rm{c}}$ {\it Physics Department and Columbia Astrophysics Laboratory,
Columbia University, New York, NY 10027, USA } \\
$^{\rm{d}}$ {\it Laboratori Nazionali di Frascati, INFN,  Frascati, Italy } \\
$^{\rm{e}}$ {\it Dipartimento di Fisica, Universit\`{a} di Milano Bicocca and INFN, 
                 Milano, Italy } \\
$^{\rm{f}}$ {\it Istituto di Cosmogeofisica, CNR, Torino, Italy} \\
$^{\rm{g}}$ {\it Laboratori Nazionali del Gran Sasso, INFN,  Assergi, Italy } \\
$^{\rm{h}}$ {\it Moscow Engineering Physics Institute, Moscow, Russia} \\
}
\end{center}

\newpage

\begin{abstract}
A high-density calorimeter, consisting of magnetized iron planes interleaved
by RPCs, as tracking and timing devices, is a good candidate for a next
generation experiment on atmospheric neutrinos. With 34~kt of mass and
in four years of data taking, this experiment will be sensitive to $\nu_{\mu}
\to \nu_x$ oscillation with $\Delta m^2 > 6 \times 10^{-5}$ and mixing near to 
maximal and fully cover the region of oscillation parameters suggested by
Super-Kamiokande results. Moreover, the experimental method will enable to
measure the oscillation parameters from the modulation of the $L/E$ spectrum
($\nu_{\mu}$ disappearance). 
For $\Delta m^2 > 3 \times 10^{-3}$ eV$^2$, this experiment 
can also establish whether the oscillation occurs into 
a tau or a sterile neutrino, by looking for an excess of muon-less
events at high energies produced by upward-going tau neutrinos
($\nu_{\tau}$ appearence).
\end{abstract}

\newpage

\pagestyle{plain}
\setcounter{page}{1}

\section{Introduction}

Recent Super-Kamiokande data \cite{superkamiokande} exhibit a zenith
angle dependent deficit of muon neutrinos ($\nu_{\mu}$) which is
inconsistent with expectations based on calculation of the atmospheric
neutrino flux.  The observation is interpreted in terms of a
two-flavour $\nu_{\mu} \to \nu_x$ oscillation with mixing $\sin^2 (2
\Theta)>0.82$ and mass-square difference $5 \times 10^{-4} \rm{eV}^2 <
\Delta m^2 <  6 \times 10 ^{-3} \rm{eV}^2$. The absence of a
corresponding deficit in the electron neutrino fluxes and data  from
reactor experiments \cite{CHOOZ} suggest that muon
neutrinos either oscillate into a tau neutrino or in a new sterile
neutrino\footnote{The sterile neutrino  hypothesis in not necessary
to explain Super-Kamiokande data, but stems from the need to reconcile
these data with other claims for neutrino  oscillation evidence.}.
This interpretation, given its relevance, needs to be tested by an
independent experiment with sensitivity to the same region of
oscillation parameters and with enough redundancy to be able to prove,
or disprove, that an observed anomaly in atmospheric neutrino fluxes
be due to neutrino oscillations.

These requirements are fulfilled by an experiment on atmospheric
neutrinos based on a large mass and high-density tracking calorimeter
\cite{98-004,curioni,NOSEX}, which has the
capability to reconstruct in  each event the $L/E$ ratio of the
neutrino path length to its energy.  As formerly suggested by P.~Picchi
and F.~Pietropaolo \cite{PioFrancesco}, $\nu_\mu$ oscillations would manifest in a
modulation of the $L/E$ spectrum, from which the oscillation
parameters can be measured.  This method ($\nu_\mu$ disappearance) has
sensitivity to $\nu_{\mu}$ oscillations with $\Delta m^2 > 6 \times
10^{-5}$~eV$^2$ and mixing near to maximal and fully covers the region
of oscillation parameters suggested by Super-Kamiokande
results. Moreover, the appearance of $\nu_{\tau}$
interactions at high energies can be searched for with the same
detector to establish whether the $\nu_\mu$ oscillation occurs into a
tau or into a sterile neutrino ($\nu_{\tau}$ appearance).

In this paper, the experimental method, the basic detector parameters
and characteristics  and its capabilities to detect oscillations of
atmospheric neutrinos are reviewed. 
With respect to previous studies, a higher sensitivity in the region of
$\Delta m^2 > 3 \times 10^{-3}$~eV$^2$ is obtained by means of a
magnetized iron detector.

\section{Experimental method}

For neutrino energies above 1.5 GeV, atmospheric neutrino fluxes are
to a good approximation up/down symmetric \cite{lipari,battistoni}.
At these energies and for $\Delta m^2 < 10^{-2}$ eV$^2$,  neutrino
oscillations would result in a modulation of the $L/E$ distribution
of upward-going neutrinos, while
downward-going neutrinos are almost unaffected by oscillations. Downward
neutrinos can therefore provide a {\it near} reference  source to
which compare the $L/E$ distribution of upward-going neutrinos ({\it
far} source). For upward neutrinos the path length $L$ is determined
by their zenith angle as $L(\theta)$, while the reference distribution
is obtained replacing the actual path length of downward neutrinos
with the mirror-distance $L'(\theta)=L(\pi - \theta)$.  The ratio
$N_{up}(L/E)/N_{down}(L'/E)$ will correspond to the survival
probability given by
\begin{equation}
\label{prob}
P(L/E)=1-\sin^2(2\Theta)\sin^2(1.27\Delta m^2L/E)
\end{equation}
with $L$ in km, $E$ in GeV, $\Delta m^2$ in eV$^2$. 
A smearing of the modulation is introduced by the finite $L/E$ resolution of
the detection method.

Some remarks are in order: i) the results obtained by this method
are to a large extent insensitive to systematics arising from calculations of
atmospheric fluxes, neutrino cross sections and detector inefficiencies; 
ii) the method does not work with neutrinos at
angles near to the horizontal ($ |\cos(\theta)| < 0.07$), since the
path lengths corresponding to a direction and its mirror-direction
are of the same order. 

If evidence of neutrino oscillation from the study of $\numu$
disappearance is obtained, a method based on $\tau$ appearance can
be used to discriminate between oscillations $\numunutau$ and $\numu
\to \nu_{sterile}$. This method consists in measuring the upward/downward
ratio of muon-less events as a function of the neutrino energy.
Oscillations of $\numu$ into $\nutau$ would in fact result in an excess of
muon-less events produced by upward neutrinos with respect to muon-less
downward. Due to threshold effects on $\tau$ production this excess would
be important at high energy. Oscillations into a sterile neutrino would
instead result in a depletion of upward muon-less events. Discrimination
between $\numunutau$ and $\numu \to \nu_{sterile}$ is thus obtained
from a study of the ratio of upward to downward muon-less events as a
function of the energy. Because this method works with the high energy
component of atmospheric neutrinos, it becomes effective for $\Delta
m^2 > 3  \times 10^{-3}$ eV$^2$.

\section{The Detector}

The outlined experimental method requires that the energy $E$ and
direction $\theta$ of the incoming neutrino be measured in each event.
The latter, in the simplest experimental approach, can be estimated
from the direction of the muon produced in the $\numu$ charged-current
interaction.  The estimate of the neutrino energy $E$ requires the
measurement of the energy of the muon and of the hadrons produced in
the interaction.  In order to make the oscillation pattern detectable,
the experimental requirement is that $L/E$ be measured with a FWHM error
smaller than half of the modulation period. This translates into
requirements on the energy  and angular resolutions of the
detector. As a general feature the resolution on $L/E$ improves at
high energies, mostly because the muon direction gives an improved
estimate of the neutrino direction. Hence, the ability to measure high
momentum muons (in the multi-GeV range), which is rather limited in
the on-going atmospheric neutrino experiments, would be particularly
rewarding.

These arguments led to consider in previous papers \cite{curioni,NOSEX} 
a large mass and high density tracking calorimeter as a suitable detector.
A large mass is necessary to provide enough neutrino interaction rate at
high energy, while the high density provided muon energy measurement by
range. Here we consider a detector of the same structure as in \cite{NOSEX},
but with the addition of a magnetic field which improves muon acceptance at
high momenta, and correspondingly efficiency at small $L/E$.

Thus, in the experiment simulation presented hereafter, a detector consisting of a
stack of 120 horizontal iron planes 8 cm thick and $15\times  30\ {\rm m^2}$
surface, interleaved by planes of sensitive elements has been considered.  
The sensitive elements (tracking devices) are housed in a 2 cm gap between the
iron planes and provide two coordinates with a pitch of 3 cm. The detector has
a total height of 12 m and a total mass exceeding 34 kt. The total surface of 
sensitive planes is 54,000 m$^2$; the number of read-out channels is 180,000. 
A magnetic induction of toroidal shape exceeds 1 T over most of the iron volume.

The elements of the sensitive planes should also enable to
identify the flight direction of the incoming neutrino. 
In fact in the $\numu$ disappearance method, if the interaction vertex is not
identified, the identification of the muon flight
direction with high efficiency and high purity is required.
This can be obtained by means of RPCs, given their time resolution
of about 2 ns \cite{RPC}.

\section{Detection of Atmospheric Neutrino Oscillations}

As outlined in section 2, detection of oscillation of atmospheric
neutrinos and measurement of their parameters will rely on two main
techniques:
\begin{itemize}
\item disappearence of events with a high-energy muon pointing upward;
\item comparison of rates upward and downward muon-less
events of high energy.
\end{itemize}

The first technique ($\numu$ disappearence) will test the hypothesis of
$\nu_{\mu}$ oscillations and measure $\Delta m^2$; the second one ($\nu_\tau$ 
appearance) will be used to discriminate between oscillations into a sterile or
a tau neutrino.  
 
A full simulation of the experimental apparatus has been implemented. 
Neutrino interactions, according to the differential flux distribution
predicted at the Gran Sasso, have been kindly provided by G.~Battistoni
and P.~Lipari. 
The GEANT package has been used for the detector simulation. 

The muon direction is obtained by a best fit procedure
to the muon track, which accounts for effects of detector resolution,
scattering and magnetic field. The muon energy is mainly determined
by range for stopping muons, by track curvature for outgoing muons.
The hadronic energy is estimated from the hit multiplicity in the
calorimeter. The detector has a coarse hadronic energy resolution
and essentially no capability of reconstructing the hadronic energy 
flow. 

\subsection{Disappearence of muon  neutrinos}

Oscillation parameters are not known {\it a priori}, therefore a unique
set of event selections and a unique analysis method have been defined
in order to make the oscillation pattern detectable for every possible
experimental outcome.  

In order to select a pure $\numu$ charged current sample,
only events with a reconstructed
track corresponding to a muon of at least 1.5 GeV were retained in the
analysis. This energy cut also insure a good up/down symmetry in absence
of oscillations. In order to reject -- in a real experiment -- the background
due to incoming muons, a further selection required the events to be either
fully contained in a fiducial volume corresponding to about 85\% of the
detector, or to have a single outgoing track (muon) with a reconstructed
range greater than 4 metres; in both samples the muon was required to hit at
least seven layers. Further selections, based on the quality of muon track fit,
and on event kinematics, were then applied in order to guarantee that the final
sample had the required $L/E$ resolution (better than 50\% FWHM) over the
whole $L/E$ range.

Altogether, these selections reduce the charged-current interaction
rate of ``unoscillated" downward muon neutrinos to about 7~kt$^{-1}
\cdot$y$^{-1}$  (20\% of the total rate of muon neutrinos above 1~GeV).
The presence of a magnetic field, which allows to include in the sample
events with an outgoing muon, increases by a factor 2 the acceptance for
$L/E$ less than 300 km/GeV.

\begin{figure}[tb]
\begin{center}
\mbox{\epsfig{file=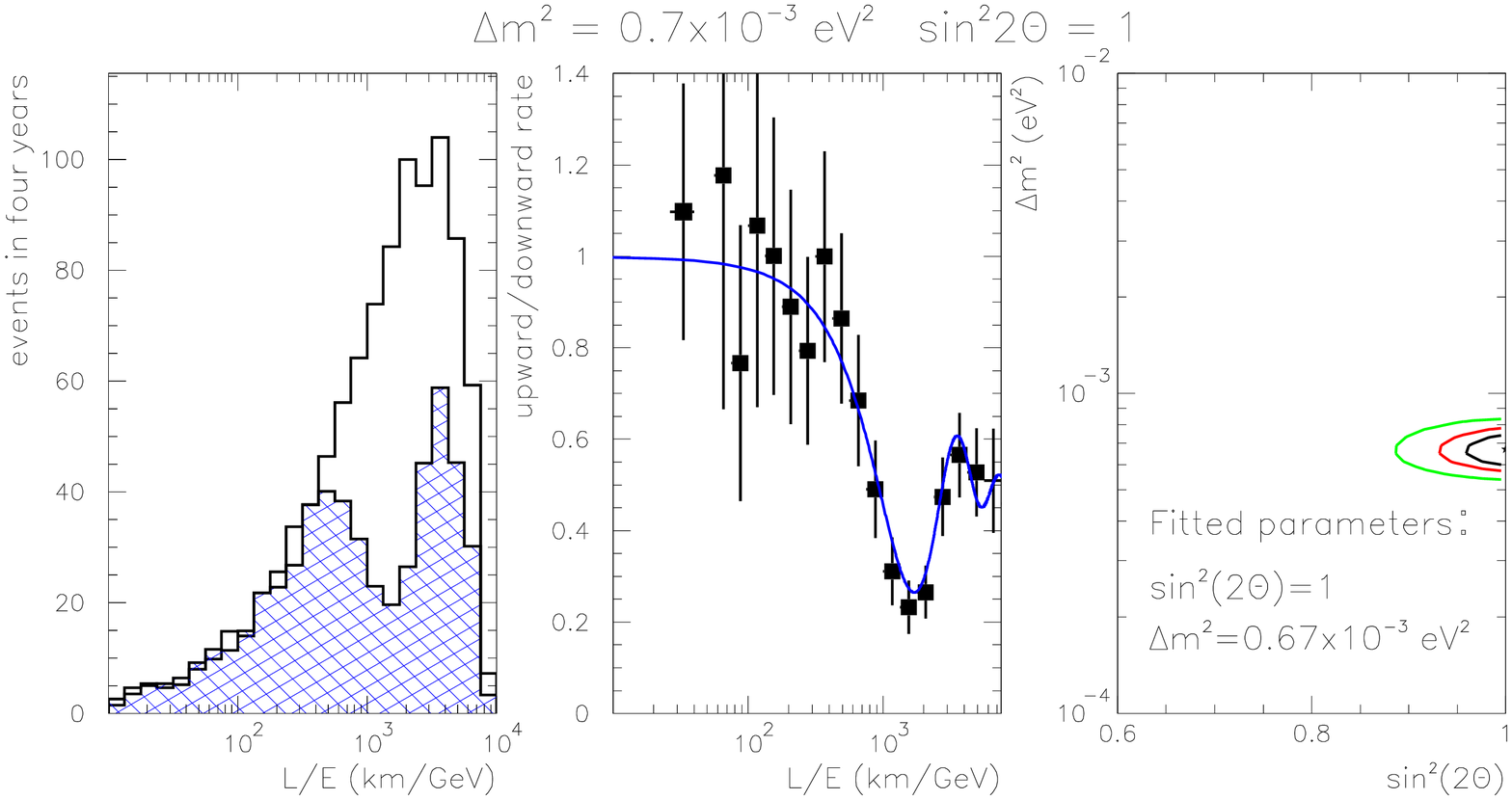,width=\linewidth}}
\mbox{\epsfig{file=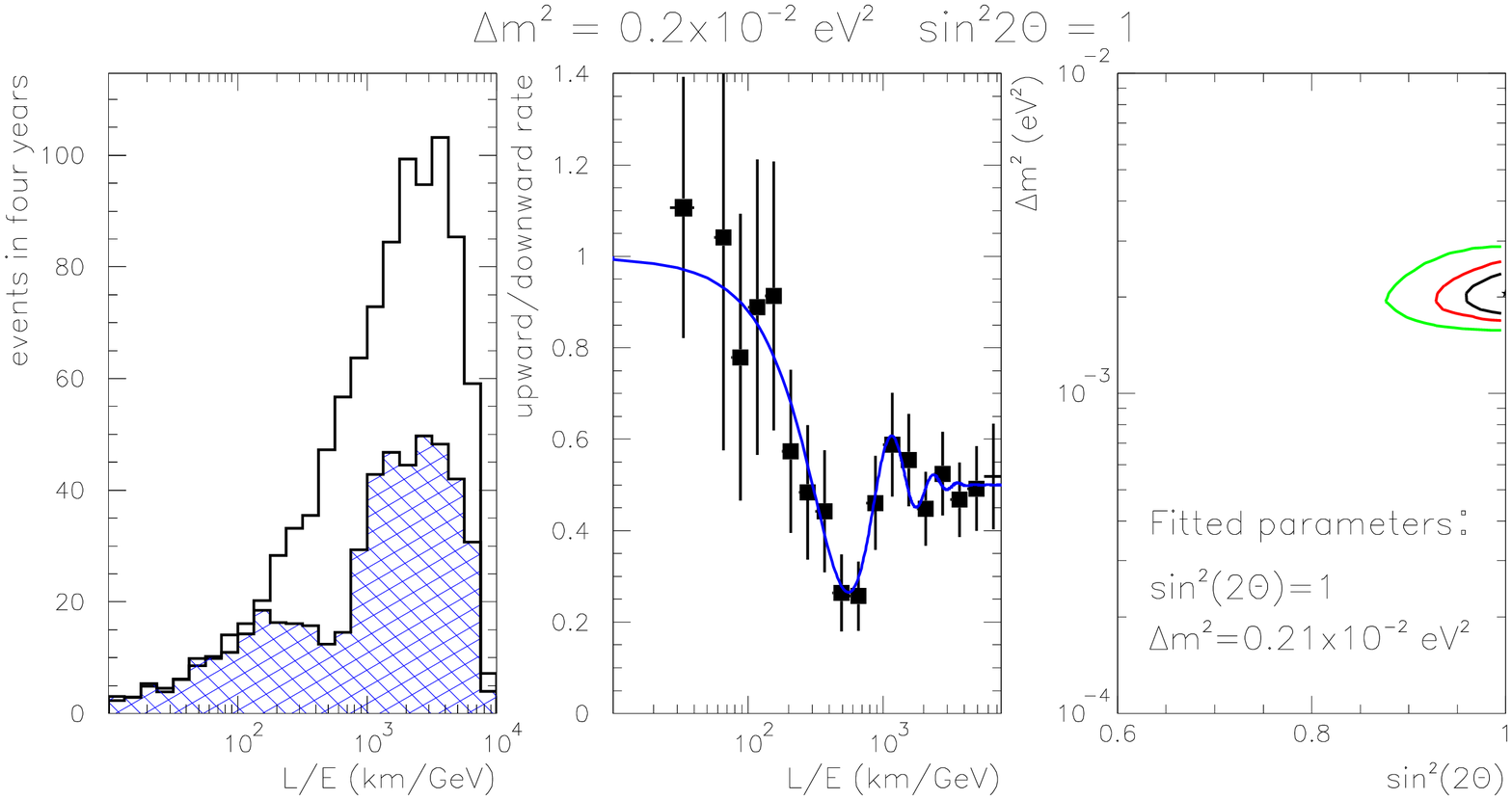,width=\linewidth}}
\end{center}
\caption{Results of the $L/E$ analysis on a simulated sample
  in presence of
  $\nu_\mu\rightarrow\nu_x$   oscillations, with parameters $\Delta
  m^2 = 7\times 10^{-4}\ eV^2$ and $\sin^2 (2\Theta)=1.0$ (top) and
  $\Delta m^2 = 2\times 10^{-3}\ eV^2$ and $\sin^2(2 \Theta) = 1.0$. 
  The figures show from left to right: %
  $L/E$ spectra for  upward muon events (hatched area) and downward 
  ones (open area); their ratio with the best-fit superimposed 
  (the first point is integrated over the first six bins) 
  and the result of the fit with the corresponding allowed regions for 
  oscillation parameters at 68\%, 90\% and 99\% C.L.. 
  Simulated statistics correspond to 25 years of data taking, rate 
  normalization, error bars and errors entering in the best
  fit procedure correspond to 4 years }
\label{fig:one}
\end{figure}

\begin{figure}[tb]
\begin{center}
\mbox{\epsfig{file=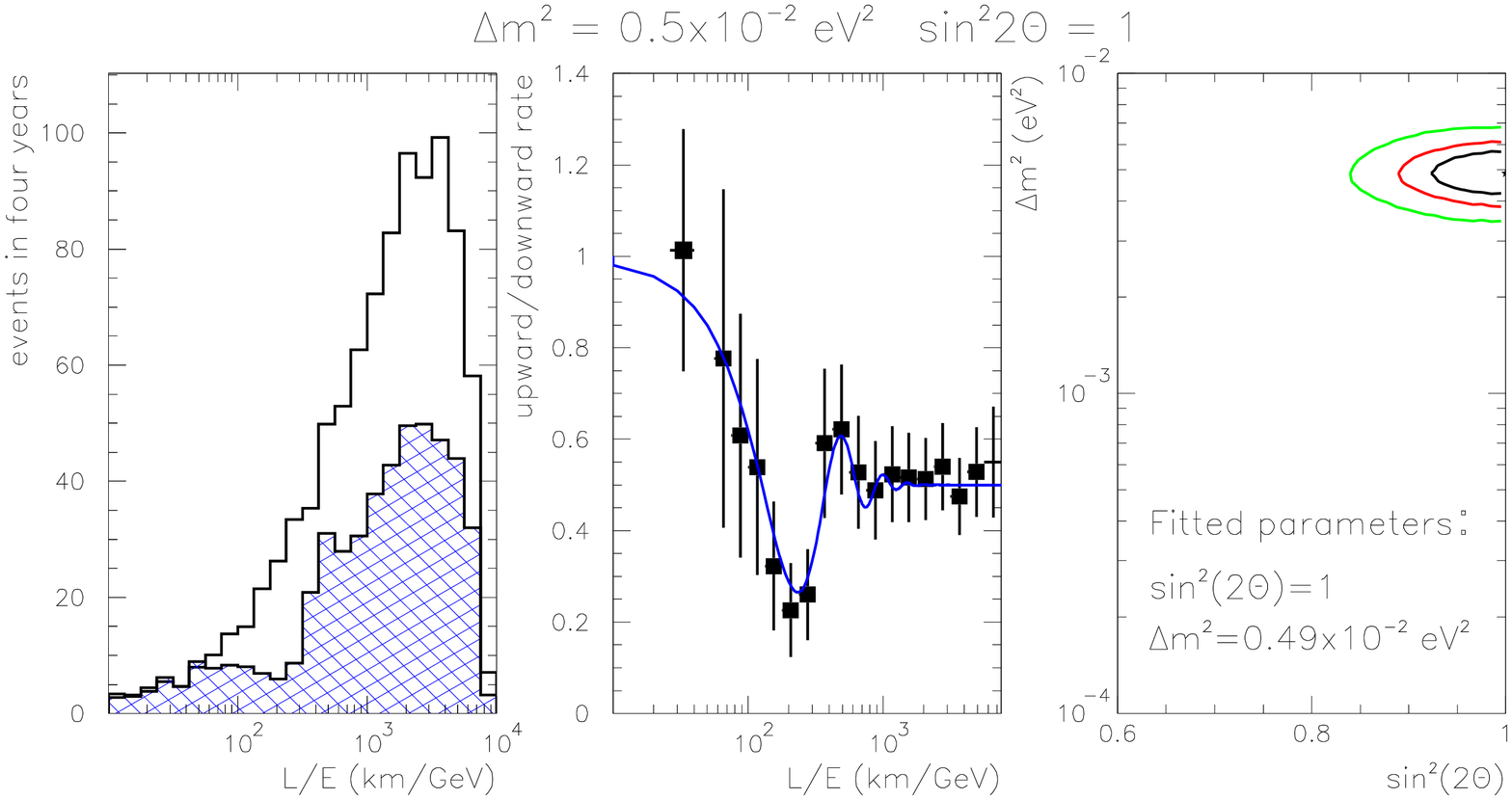,width=\linewidth}}
\mbox{\epsfig{file=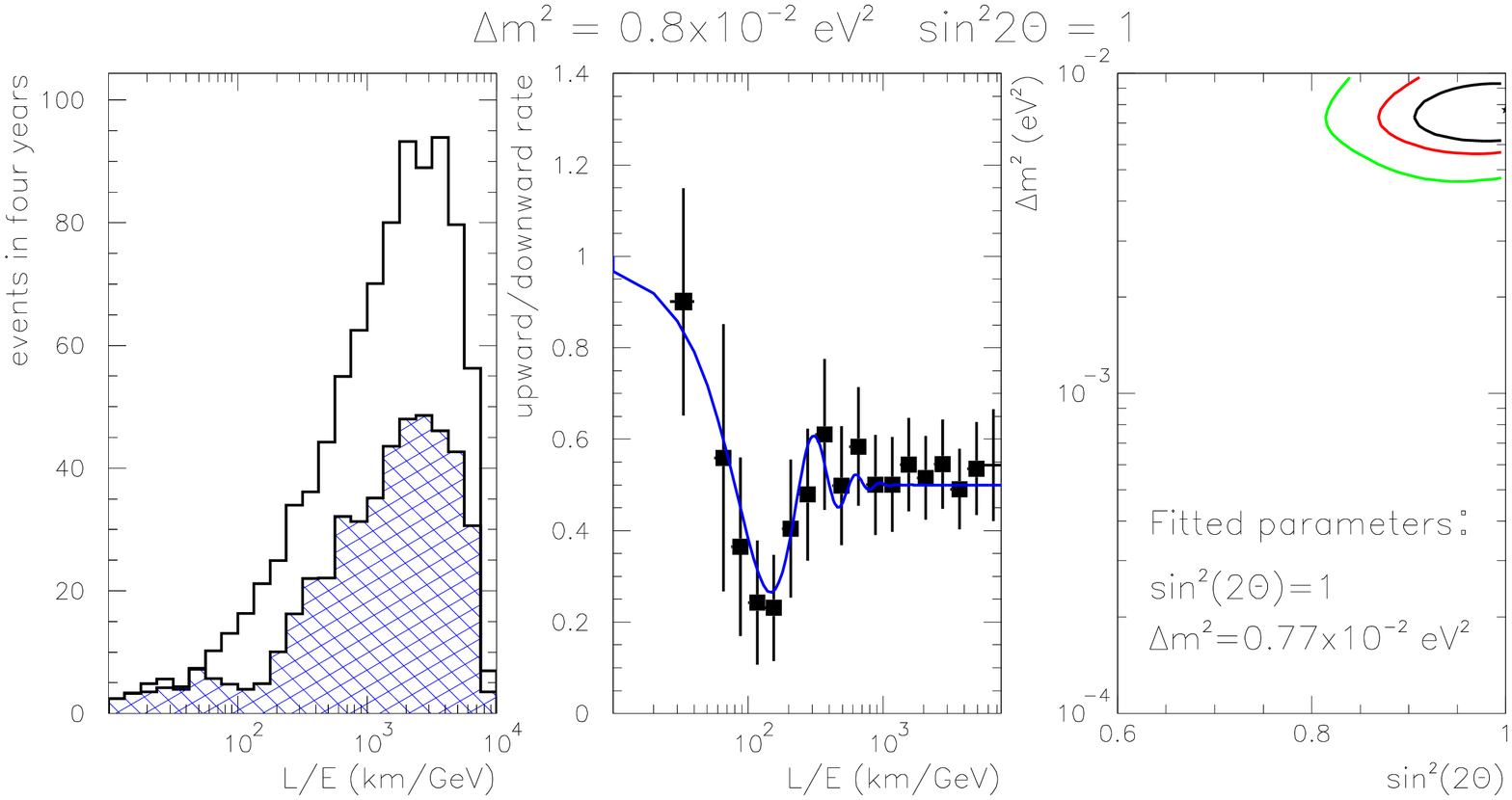,width=\linewidth}}
\end{center}
\caption{As Fig.~\ref{fig:one}, for $\Delta m^2 = 5 \times 10^{-3}\
  eV^2$ and $\sin^2(2 \Theta) = 1.$ (top) and 
  $\Delta m^2 = 8 \times 10^{-3}\
  eV^2$ and $\sin^2(2 \Theta) = 1.$ (bottom)}
\label{fig:two}
\end{figure}

The $L/E$ distributions obtained with the outlined selections
are shown in Fig.~\ref{fig:one} and Fig.~\ref{fig:two}
for $\Delta m^2$ ranging from $7\times 10^{-4}$ to
$8\times 10^{-3}~\rm{eV}^2$ and maximum mixing.
The figures also show the discovery potential (allowed
regions of the oscillation parameter space) of the experiment after
four years of exposure, as derived from a fit to the $L/E$ spectra of
a predictive curve folded with the detector resolution.  

\begin{figure}[tb]
\begin{center}
\mbox{\epsfig{file=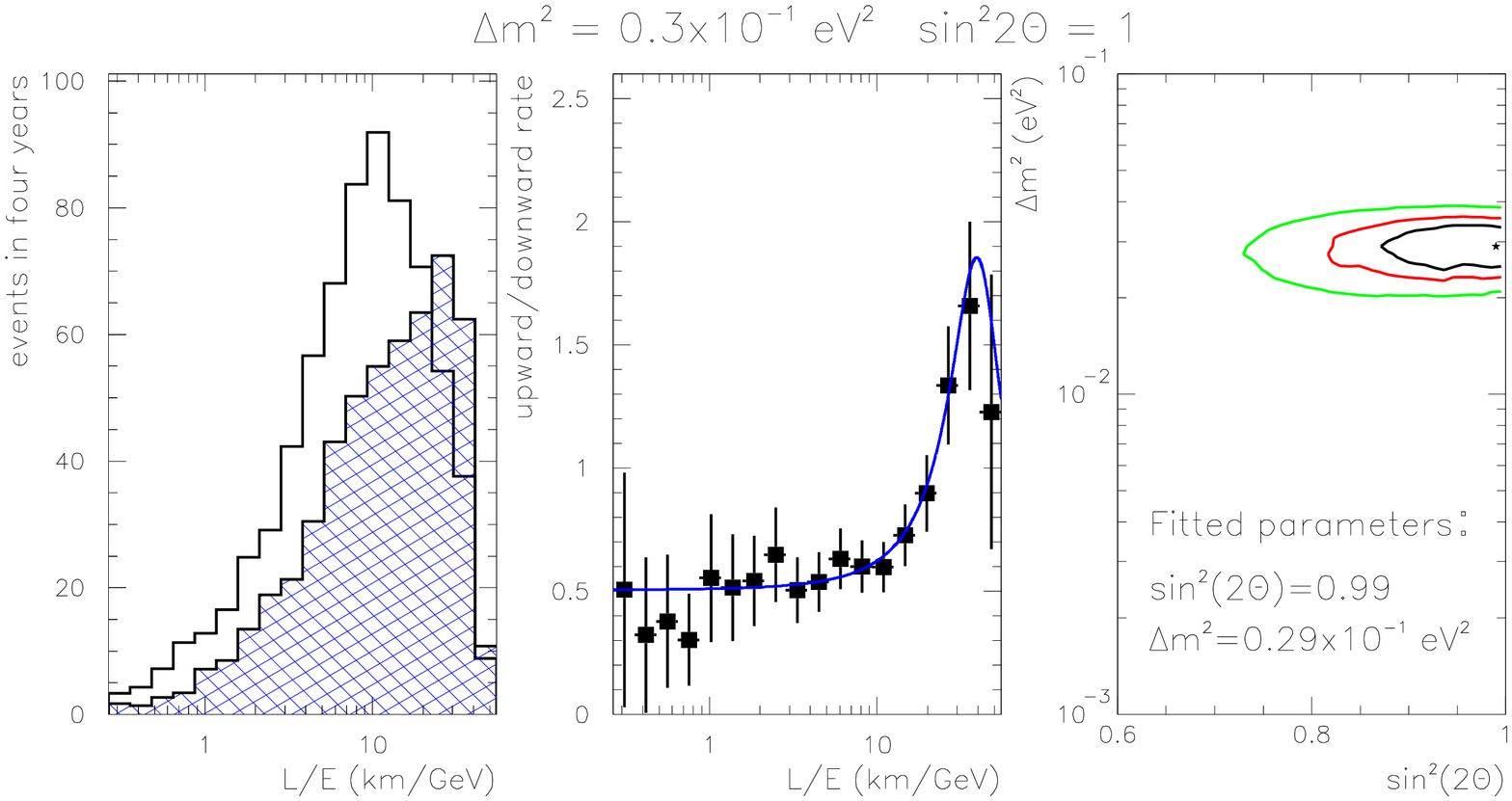,width=\linewidth}}
\end{center}
\caption{
As Fig.~\ref{fig:one}, for $\Delta m^2 = 3 \times 10^{-2}\
  eV^2$ and $\sin^2(2 \Theta) = 1.$ As explained in the text,
  an {\it inverse} oscillation pattern appears  for large values
  of  $\Delta m^2$
   }
\label{inversa}

\begin{center}
\mbox{\epsfig{file=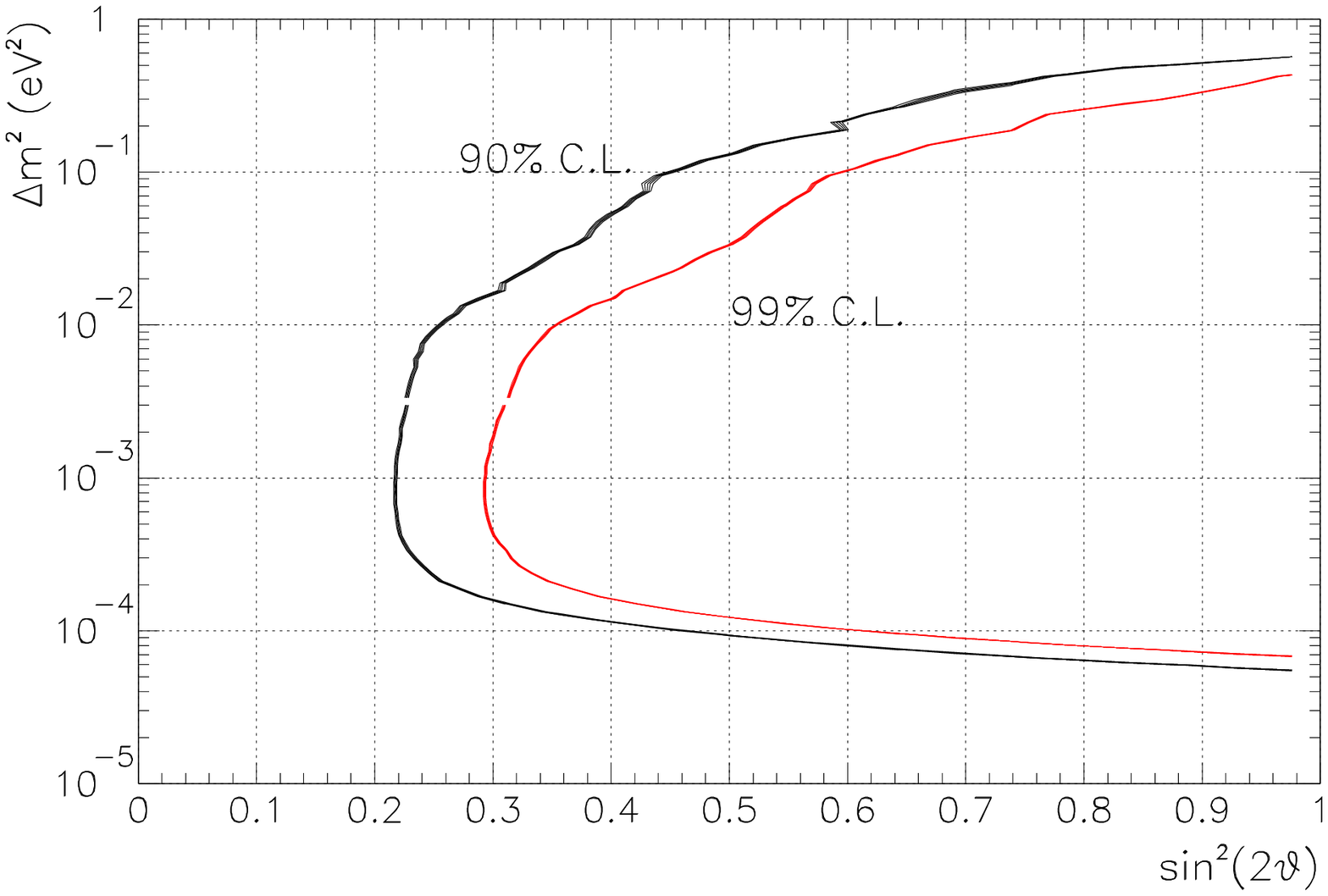,width=\linewidth}}
\end{center}
\caption{ Exclusion curves at 90\% and 99\% C.L. after three years
  of data taking assuming no oscillations.}
\label{sensibilita}
\end{figure}

We also notice that if  $\Delta m^2$ were larger than a few
$10^{-2}$ eV$^2$, upward neutrinos -- at large $L/E$ -- would be in 
complete oscillation,
while the oscillation pattern would become detectable in the downward sample.
In this limit, a mirror distance $L'(\theta)=L(\pi - \theta)$ can be assigned
to upward neutrinos, which can be used as a reference $L/E$ distribution
for downward neutrinos. In this case, due to the uncertain estimate of the
neutrino path-length for downgoing neutrinos related to our ignorance of
their production height in the atmosphere, there would be some model
dependence in the determination of oscillation paramenters. Nonetheless,
the observation of an oscillation pattern would still firmly test the
oscillation hypothesis.
An example of the $L/E$ distribution and results obtained with this
analysis are shown in Fig.~\ref{inversa}.

In absence of neutrino oscillations, these arguments can be used to exclude a
region of oscillation parameters. The exclusion limits at 90\% and 99\% C.L.
that this experiment will be able to set after an exposure of three years are
shown in Fig.~\ref{sensibilita}. A 2\% systematic uncertainty in the knowledge of the up/down
ratio of atmospheric neutrino fluxes has been assumed. 

\subsection{Appearance of tau neutrinos}

If evidence of $\numu$ disappearance is observed, for $\Delta m^2 
> 3 \times 10^{-3} \rm{eV}^2$, the appearence of tau neutrino
charged-current interactions can be searched for to distinguish
between $\nu_\mu\rightarrow \nu_\tau$ and $\nu_\mu\rightarrow
\nu_{sterile}$ oscillation. As discussed in Section 2, this method
consists in measuring the upward/downward ratio of muon-less events,
as a function of the visible energy. Charged-current $\nu_\tau$ interactions 
would in fact result in an excess of muon-less events in the upward
sample at high energies, due to the large tau branching ratio into
muon-less channels ($BR \simeq 0.8$). Moreover, because of threshold 
effects on tau production, events of large visible energy must be
selected, in order to enhance the relative $\nu_\tau$ contribution to
the muon-less event sample.  

In the analysis, an event has been considered a muon-less candidate 
if it did not contain non-interacting tracks longer than $1\ \rm{m}$
(equivalent to $0.9\ \rm{GeV}$ for a m.i.p.). An estimate of the
visible energy has been obtained from the hit multiplicity  in the two
views; the up/down direction has been derived from the shape of the
hadronic shower development. The analysis of simulated data has been
performed first through visual scanning to optimise the up/down
discrimination efficiency, then in an automated way where the best
selection cuts have been implemented. Events with an ambiguous
determination of the neutrino flight direction have been discarded. 

These selections also retain neutral-current neutrino interactions, 
$\nu_e$ charged-current interactions and $\nu_\mu$ charged-current
events with a soft muon. However, the rejection of $\nu_\mu$-CC events is
effective at high energy, because of the cut of muon with energy larger
than 0.9 GeV (due to the flat y distribution of the CC interaction). 
The $\nu_e$-CC background is mostly degraded to low visible energy,
due to the coarse digital sampling of the detector which  filters off the
electro-magnetic component of the interaction. As a consequence the
visible energy is only due to the residual hadronic component, as in
the case of neutral current events. 

\begin{figure}[tb]
\begin{center}
\mbox{\epsfig{file=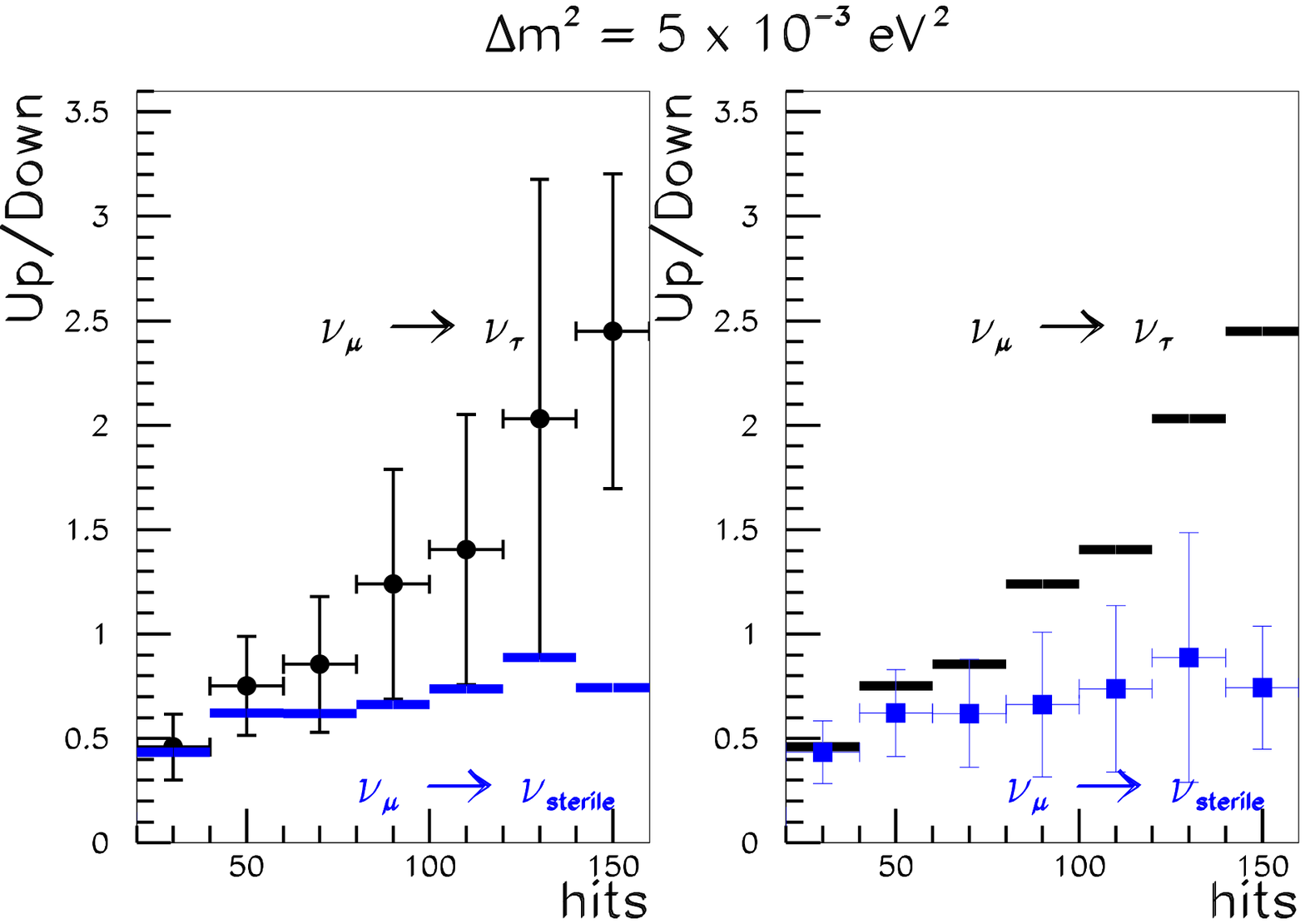,width=\linewidth}}
\end{center}
\caption{Up/down differential ratios of muon-less events as a function
  of the number of hits, for $\Delta m^2 = 5\times 10^{-3}\rm\ eV^2$
  and maximal mixing. The result of the simulation for $\numu \to
  \nutau$ is compared to the expectations  for $\numu \to
  \nu_{sterile}$ oscillations (left) and vice versa (right). Events
  have been generated with high statistics, error bars correspond to
  three years of data taking. The rightmost bin also integrates the
  contribution of events with hit multiplicity larger than 160.} 
\label{tauste}
\end{figure}

Fig.~\ref{tauste} shows the differential up/down ratio as a function of the hit
multiplicity  in the calorimeter, for $\Delta m^2 = 5 \times
10^{-3}$ eV$^2$. 
In the $\nu_\mu\rightarrow\nu_\tau$ case there is an excess of muon-less
events with high visible energy from the bottom hemisphere due to the
tau decay into muon-less channels that produce neutral current like events; 
in the $\nu_\mu\rightarrow\nu_{sterile}$ case there is a lack of neutral
currents from the bottom hemisphere at all visible energies, for the sterile
neutrino does not interact. 
For $\Delta m^2 = 5 \times 10^{-3}$ eV$^2$ and in three years of data taking,
the two alternative hypothesis can be discriminated at the 90\% C.L. 
with a rejection power of $10^{-2}$, corresponding to a separation 
of about $3 \sigma$. A similar separation is obtained for larger values 
of $\Delta m^2$. 

\section{Conclusions}

A high density calorimeter of 34 kt with rough sampling and good tracking
capability is a good candidate for a next generation experiment on atmospheric
neutrinos. The detector has an estimated cost of abut 20 MEuro and can be
built in a short delay. 

In three years of data taking, this experiment will be sensitive to
$\nu_{\mu} \to \nu_x$ oscillation with $\Delta m^2 > 6 \times 10^{-5}$
and mixing near to maximal and fully cover the region of oscillation
parameters suggested by Super-Kamiokande results. Moreover, the
experimental method will enable to measure the oscillation parameters
from the modulation of the $L/E$ spectrum ($\nu_{\mu}$ disappearance)
and to establish whether the oscillation occurs into a tau or into a
sterile neutrino ($\nu_{\tau}$ appearance).  

The major improvement with respect to Super-Kamiokande relies on the 
exploitation of the high energy component of the atmospheric muon
neutrino spectrum, which reflects in a better $L/E$ resolution in the
range around $10^3$~{km/GeV}. 

This experiment, completely devoted to the atmospheric neutrino study,
is  complementary to those designed for long base line neutrino
detection with artificial neutrino beams. In fact, in the energy range
considered here, atmospheric neutrinos are still an unexploited source
of potential  discovery, since they cover a wider $L/E$ range than
long baseline beams \cite{battistoni}.

\section*{Acknowledgements}

We gratefully acknowledge F.~Pietropaolo, for his invaluable 
contribution to this work, and P.~Lipari and G.~Battistoni for 
providing us the neutrino events, without which this work would not 
have been possible.

\pagebreak

\newpage

\newpage

\begin{thebibliography}{000}

\bibitem{superkamiokande}
Super-Kamiokande Collaboration, Y. Fukuda et al.,
\Journal{\PRL}{81}{1562-1567}{1998} \\
Super-Kamiokande Collaboration, Y. Fukuda et al.,
\Journal{\PLB}{436}{33}{1998} \\
Super-Kamiokande Collaboration, Y. Fukuda et al.,
\Journal{\PLB}{433}{9}{1998}

\bibitem{CHOOZ}
M. Apollonio et al., \Journal{\PLB}{420}{397-404}{1998}

\bibitem{98-004} G.~Mannocchi et al., CERN/OPEN-98-004

\bibitem{curioni} A.~Curioni et al., hep-ph/9805249
\bibitem{NOSEX} M.~Aglietta et al., CERN/SPSC 98-28 SPSC?M615, Oct.~1998

\bibitem{PioFrancesco} P.~Picchi and F.~Pietropaolo, ``Atmospheric Neutrino Oscillations Experiments'',ICGF RAP. INT. 344/1997, Torino 1997, (CERN preprint
SCAN--9710037).


\bibitem{lipari} P.~Lipari, T.~K.~Gaisser and T.~Stanev, astro-ph/9803093
\bibitem{battistoni}  G.~Battistoni and P.~Lipari, hep-ph/9807475

\bibitem{RPC} G.~Bencivenni et al., \Journal{\NIMA}{345}{456}{1994}

\end{thebibliography}
\end{document}